\documentclass[12pt,twoside,english]{article}
\usepackage[T1]{fontenc}
\usepackage[latin1]{inputenc}
\usepackage{amsbsy}
\usepackage{amssymb}
\usepackage{graphicx}
\usepackage{setspace}


\usepackage{babel}
\begin{document}
\title{Time-magnitude correlations and time variation of the Gutenberg-Richter
parameter in foreshock sequences}
\author{{\normalsize{}B. F. Apostol$^{a*}$ and L. C. Cune$^{b}$ }\\
{\normalsize{}$^{a}$Institute of Earth's Physics, $^{b}$ Institute
of Physics and Nuclear Engineering}\\
{\normalsize{}Magurele-Bucharest MG-6, POBox MG-35, Romania }\\
{\normalsize{}$^{*}$corresponding author, email: afelix@theory.nipne.ro}}
\date{{}}

\maketitle
\relax
\begin{abstract}
The time dependence of the parameter of the Gutenberg-Richter (GR)
magnitude distribution is computed for foreshock sequences of earthquakes,
correlated with the main shock, by using the geometric-growth model
of earthquake focus, the magnitude distribution of correlated earthquakes
and the time-magnitude correlations, derived recently. It is shown
that this parameter decreases in time in the foreshock sequence, from
the background values down to the main shock. If correlations are
present, this time dependence and the time-magnitude correlations
provide a tool of monitoring the foreshock seismic activity. We discuss
the relevance of such an analysis for the occurrence moment and the
magnitude of a main shock. The discussion is applied to a few Vrancea
main shocks and the precursory seismic activity of the l'Aquila earthquake.
The limitations of such an analysis are discussed.
\end{abstract}
\relax


\section{Introduction}

Recently, Gulia and Wiemer (2019) suggested that the difference between
the parameters ($\beta$) of the Gutenberg-Richter (GR) magnitude
distribution of the aftershocks and the foreshocks can be used to
estimate the occurrence of main shocks. Accompanying earthquake sequences
have been analyzed by these authors for the Amatrice-Norcia earthquakes
(24 August 2016, magnitude $6.2$; 30 October 2016, magnitude $6.6$)
and the Kumamoto earthquakes (15 April 2016, magnitude $6.5$ and
$7.3$). They found that the foreshock parameter $\beta$ is lower
than the background value (\emph{e.g.}, by $10\%$), while the aftershock
parameter is higher than the background value (\emph{e.g.}, by $20\%$).
A similar decrease in the parameter $\beta$ has been reported for
the foreshocks of the L'Aquila earthquake (6 April 2009, magnitude
$6.3$) by Gulia et al. (2016) and the Colfiorito, Umbria-Marche,
earthquake (26 September 1997, magnitude $6$) by De Santis et al.
(2011). The analysis method employed by Gulia and Wiemer (2019) was
recently questioned (Dascher-Cousineau et al. 2020, 2021; see also
Gulia and Wiemer 2021). We put forward in this paper a method, based
on a short-term analysis of the foreshocks, which may be relevant
for estimating the occurrence time and the magnitude of the main shocks.
The method is based on the correlations which may exist between foreshocks
and the main shock.

The standard GR magnitude distribution is $P(M)=\beta e^{-\beta M}$
(moment magnitude $M$), where the parameter $\beta$ varies in the
range $1.15$ to $3.45$ (in decimal basis $0.5$ to $1.5$); the
mean value $\beta=2.3$ (in decimal basis $\beta=1$) is usualy accepted
as a reference value (Stein and Wysession 2003; Udias 1999; Lay and
Wallace 1995; Frohlich and Davis 1993). It has been shown (Apostol
2006) that $\beta=br$, where $b=3.45$ (in decimal basis $3/2$)
and $r$ is a parameter characterizing the earthquake focus (see Appendix).
The parameter $r$ is a statistical parameter which reflects mainly
the average number of dimensions of the focus. We expect the parameter
$r$ to vary between $r=1/3$ and $r=1$, with a mean value $r=2/3$
($\beta=2.3$). The standard cumulative (excedence) GR distribution
(earthquakes with magnitude greater than $M$) is $P_{ex}(M)=e^{-\beta M}$;
it is used in its logarithmic form $\ln N(M)=\ln N(0)-\beta M$, where
$N(M)$ is the number of earthquakes with magnitude greater than $M$. 

According to these standard formulae, an increase in $\beta$ indicates
the occurrence of more small-magnitude earthquakes, which may appear
in the aftershock region, while a decrease in $\beta$ indicates,
comparatively, more greater-magnitude earthquakes. A decrease in $\beta$
in the foreshock region has been reported in many instances (see,
\emph{e.g.}, Gulia et al. 2016 and References therein), as well as
an increase in the aftershock region (Gulia et al. 2018). In principle,
a statistical description of the accompanying seismic activity implies
a symmetric distribution in the foreshock-aftershock regions. However,
after a main shock the condition of the seismic region may change
appreciably, such that it is difficult to view the foreshocks and
the aftershocks as members of the same statistical ensemble. 

\section{Correlations}

Earthquakes which occur closely in time and space, like the earthquake
sequence accompanying a main shock, may be correlated with the main
shock (see Appendix). The magnitude distribution of the correlated
earthquakes differs from the standard Gutenberg-Richter distribution
discussed above (Apostol 2021). Judged by their time-dependence shape,
the first part of the foreshock distribution indicated by Gulia and
Wiemer (2019) may exhibit correlations, but correlations cannot be
definitely assessed in the aftershocks distribution; a change in the
seismicity conditions may be present for aftershocks. We discuss below
a possible relevance of a correlated foreshock sequence for the occurrence
of a main shock. 
\begin{figure}
\begin{centering}
\includegraphics[clip,scale=0.3]{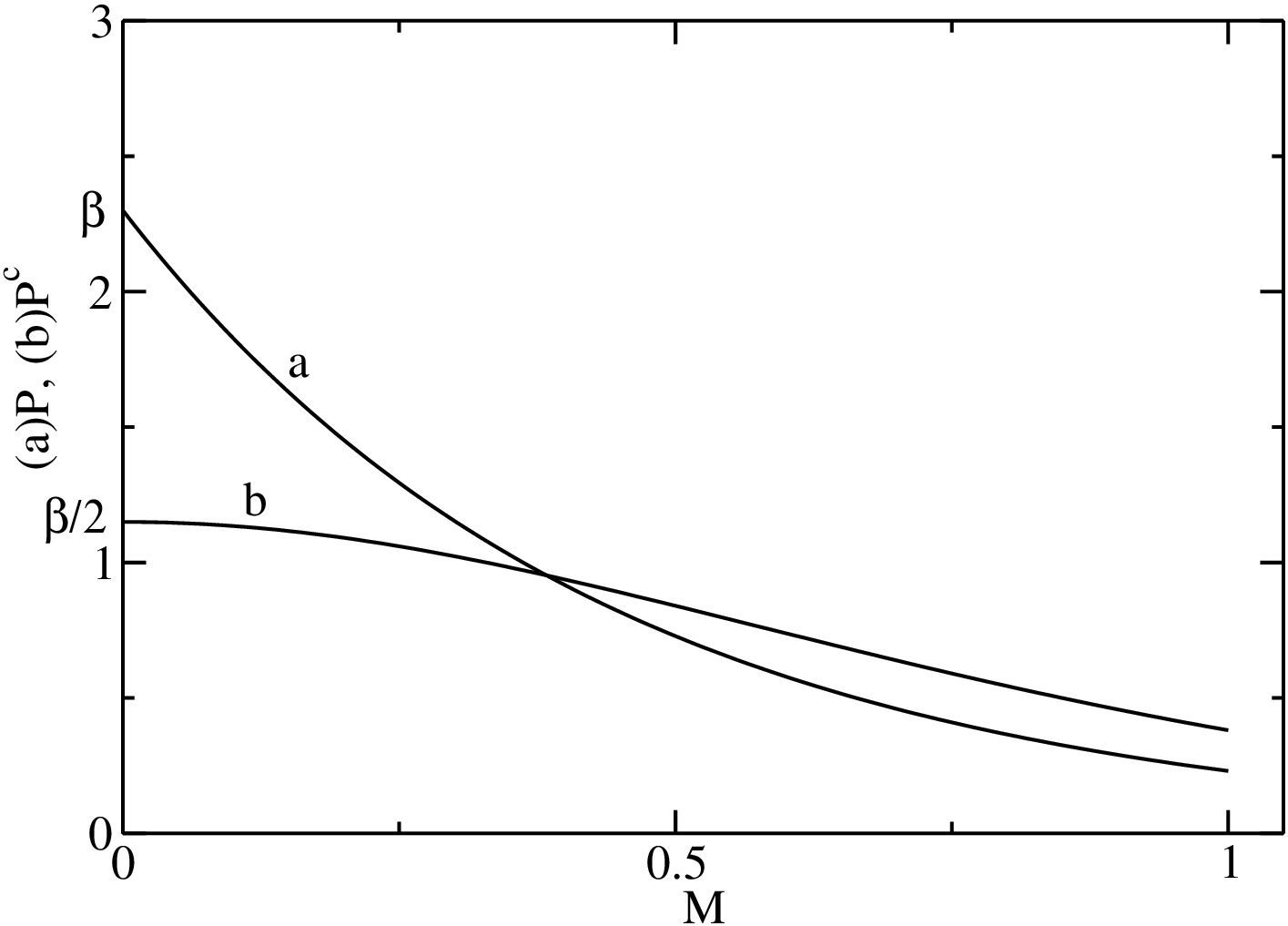}
\par\end{centering}
\caption{The standard GR distribution $P=\beta e^{-\beta M}$ (curve a) compared
to the correlation-modified GR distribution $P^{c}$, Eq. (\ref{1})
(curve b).\label{Fig.1}}

\end{figure}

The earthquake correlations, as identified in Apostol (2021), are
dynamical, purely statistical and time-magnitude correlations. The
dynamical correlations (called also \textquotedbl causal\textquotedbl{}
correlations) imply a sharing of the accumulation time; they may also
be viewed as temporal correlations. These correlations lead to modified
statistical distributions, as the modified GR distribution given below.
They may appear by a static or a dyamical stress, a change in the
seismicity conditions of the focal region, a triggering mechanism,
etc. Mathematical conditions (constraints) imposed on the statistical
variables give rise to purely statistical correlations. Time-magnitude
correlations, which are discussed in this paper, are particular dynamical
correlations arising from the non-linearity of the law of energy accumulation
in the focus. This law allows an energy sharing between two (or more)
earthquakes, which makes an earthquake to depend on the other, \emph{i.e.}
it generates a correlation between these earthquakes (see Appendix). 

\begin{figure}
\begin{centering}
\includegraphics[clip,scale=0.3]{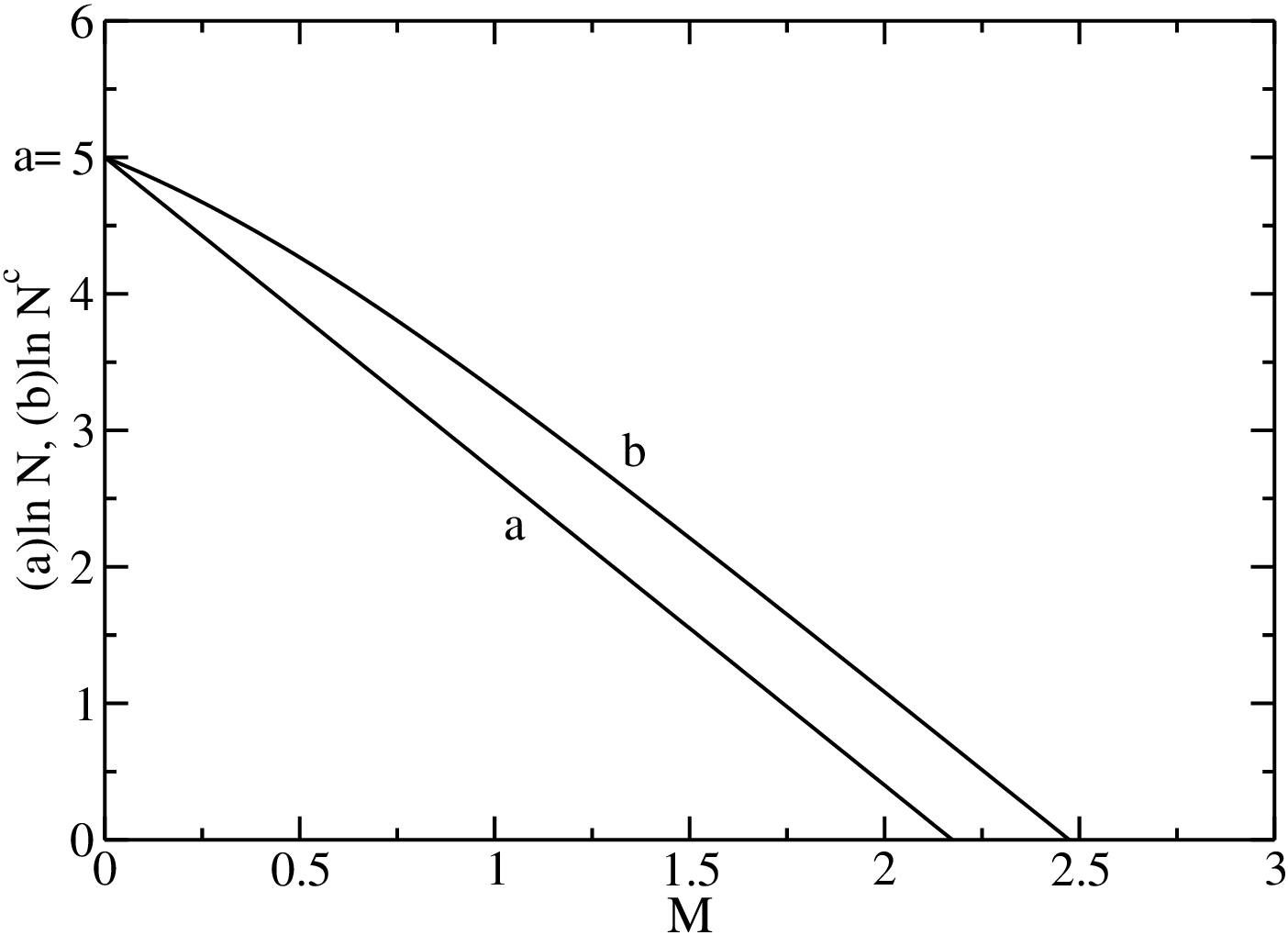}
\par\end{centering}
\caption{The standard cumulative GR distribution $\ln N=\ln N(0)-\beta M$
(curve a) compared to the correlation-modified cumulative GR distribution
$\ln N^{c}$, Eq. (\ref{3}) (curve b) for $\beta=2.3$ and an arbitrary
value $\ln N(0)=5$.\label{Fig.2}}

\end{figure}

The correlation-modified magnitude distribution (modified GR distribution,
Apostol 2021 ) is 
\begin{equation}
P^{c}(M)=\beta e^{-\beta M}\frac{2}{(1+e^{-\beta M})^{2}}\,\,;\label{1}
\end{equation}
without other specifications, this distribution includes the so-called
dynamical correlations, which affect mainly the small-magnitude earthquakes.
We expect such correlations to be present in foreshock sequences.
From Eq. (\ref{1}) we get the correlation-modified cumulative distribution
\begin{equation}
P_{ex}^{c}(M)=e^{-\beta M}\frac{2}{1+e^{-\beta M}}\,\,.\label{2}
\end{equation}
 The logarithmic form of this distribution
\begin{equation}
\ln N^{c}(M)=\ln N(0)+\ln2-\ln\left(1+e^{\beta M}\right)\label{3}
\end{equation}
 should be compared to the standard logarithmic form 
\begin{equation}
\ln N(M)=\ln N(0)-\beta M\,\,.\label{4}
\end{equation}
We can see that the modified GR distributions (Eqs. (\ref{1}) and
(\ref{2})) differ from the standard GR distributions, as shown in
Figs. \ref{Fig.1} and \ref{Fig.2}. It seems that such a qualitative
difference has been found for southern California earthquakes recorded
between 1945-1985 and 1986-1992 (Jones 1994). The difference arises
mainly in the small-magnitude region $M\lesssim1$, where the distributions
are flattened. For instance, in this region the parameter $\beta$
of the cumulative distribution tends to $\beta/2$, according to Eqs.
(\ref{1}) and (\ref{2}) (see Appendix). This deviation, known as
the roll-off effect (Bhattacharya et al. 2009; Pelletier 2000), is
assigned usually to an insufficient determination of the small-magnitude
data. We can see that it may be due to correlations, at least partially.
For large magnitudes the logarithmic cumulative distribution is shifted
upwards by $\ln2$ (Eq. (\ref{3})), while its slope is very close
to the slope of the standard cumulative GR distribution ($\beta$).

The correlation-modified cumulative distribution given by Eq. (\ref{2})
can be used to identify a correlated sequence of foreshocks. We consider
a seismic region with a background of earthquakes (regular earthquakes)
extended over a long period of time $T$, interrupted from time to
time by (rare) big seismic events. We may assume that some of these
large earthquakes are main shocks in accompanying sequences of foreshocks
(and aftershocks), correlated with the main shock. For moderate and
large magnitudes we may fit the seismic activity by the standard cumulative
GR distribution given by Eq. $\ln N(M)=\ln N(0)-\beta M$. Usually,
such fits are done by using a small-magnitude cutoff, such that the
slope of the distribution ($\beta$) is not affected by correlated
small-magnitude earthquakes (difficulties in determining the $\beta$-parameter
by finite sets of data and the related completeness magnitude are
discussed recently by Marzocchi et al. 2020 and Lombardi 2021). A
proper fitting of the full (modified) GR distribution given by Eq.
(\ref{3}) leads to very small differences in the parameter $\beta$.
It is convenient to introduce the parameter $t_{0}=T/N(0)$; its inverse
is a seismicity rate. Due to the small-magnitude cutoff, this seismicity-rate
parameter becomes a fitting parameter (Apostol 2021). The standard
GR cumulative distribution reads
\begin{equation}
\ln\left[N(M)/T\right]=-\ln t_{0}-\beta M\,\,.\label{5}
\end{equation}
By fitting this law to the empirical data we get the parameters $\beta$
(and $r$) and $t_{0}$. For instance, such a fit, done for a set
of $3640$ earthquakes with magnitude $M\geq3$ which occurred in
Vrancea during $1981-2018$, leads to $-\ln t_{0}=11.32$ ($t_{0}$
measured in years) and $\beta=2.26$ ($r=0.65$), with an estimated
$15\%$ error. We note that the value $\beta=2.26$ is close to the
reference value given above ($2.3$). (The data for Vrancea have been
taken from the Romanian Earthquake Catalog 2018, http://www.infp.ro/data/romplus.txt.
A completeness magnitude $M=2.2$ to $M=2.8$ is usually accepted
(Enescu et al. 2008 and References therein); a more conservative figure
would be $M=3$. The magnitude average error is $\Delta M=0.1$).
A similar fit, with slightly modified parameters, is valid for $8455$
Vrancea earthquakes with magnitude $M\geq2$ (period $1980-2019$).
This way, we get the parameters of the background seismic activity
for Vrancea ($\beta$, $r$, $t_{0}$).

\section{Time-magnitude formula}

Let us assume now that we are in the proximity of a main shock with
magnitude $M_{0}$, at time $\tau$ until its occurence, and we monitor
the sequence of correlated foreshocks. It was shown (Apostol 2021)
that the magnitudes of the (correlated) foreshocks $M$ ($<M_{0}$)
are related to the time $\tau$ by 
\begin{equation}
M\simeq\frac{1}{b}\ln(\tau/\tau_{0})\,\,\,,\label{6}
\end{equation}
where 
\begin{equation}
\tau_{0}=rt_{0}e^{-b(1-r)M_{0}}\label{7}
\end{equation}
 is a cutoff time, which depends on the magnitude of the main shock,
the seismicity-rate parameter $t_{0}$ and the parameter $r=\beta/b$
(see Appendix). The parameters $t_{0}$ and $r$ are provided by the
analysis of the background seismic activity. The small threshold time
$\tau_{0}$ corresponds to a very short quiescence time (Ogata and
Tsuruoka 2016) before the occurrence of the main shock. In addition,
the time $\tau$ should be cut off by an upper threshold, corresponding
to the magnitude of the main shock ($\tau<\tau_{0}e^{bM_{0}}$). We
limit ourselves to small and moderate magnitudes $M$ in the accompanying
seismic activity, such that the magnitude of the main shock may be
viewed as being sufficiently large (in this respect, the so-called
purely statistical correlations discussed by Apostol (2021), are not
included). Eq. (\ref{6}) is derived by analyzing the time-magnitude
correlations predicted by the geometric-growth model of earthquake
focus (Apostol 2006, see Appendix). According to this model the accumulation
time of an earthquake with energy $E$ is $t=t_{0}(E/E_{0})^{r}=t_{0}e^{\beta M}$,
where $E_{0}$ is a cutoff energy. By means of this model, Bath's
law is derived and the occurrence time of the Bath partner is calculated,
as well as the cumulative magnitude distribution of the accompanying
seismic activity.
\begin{figure}
\begin{centering}
\includegraphics[clip,scale=0.3]{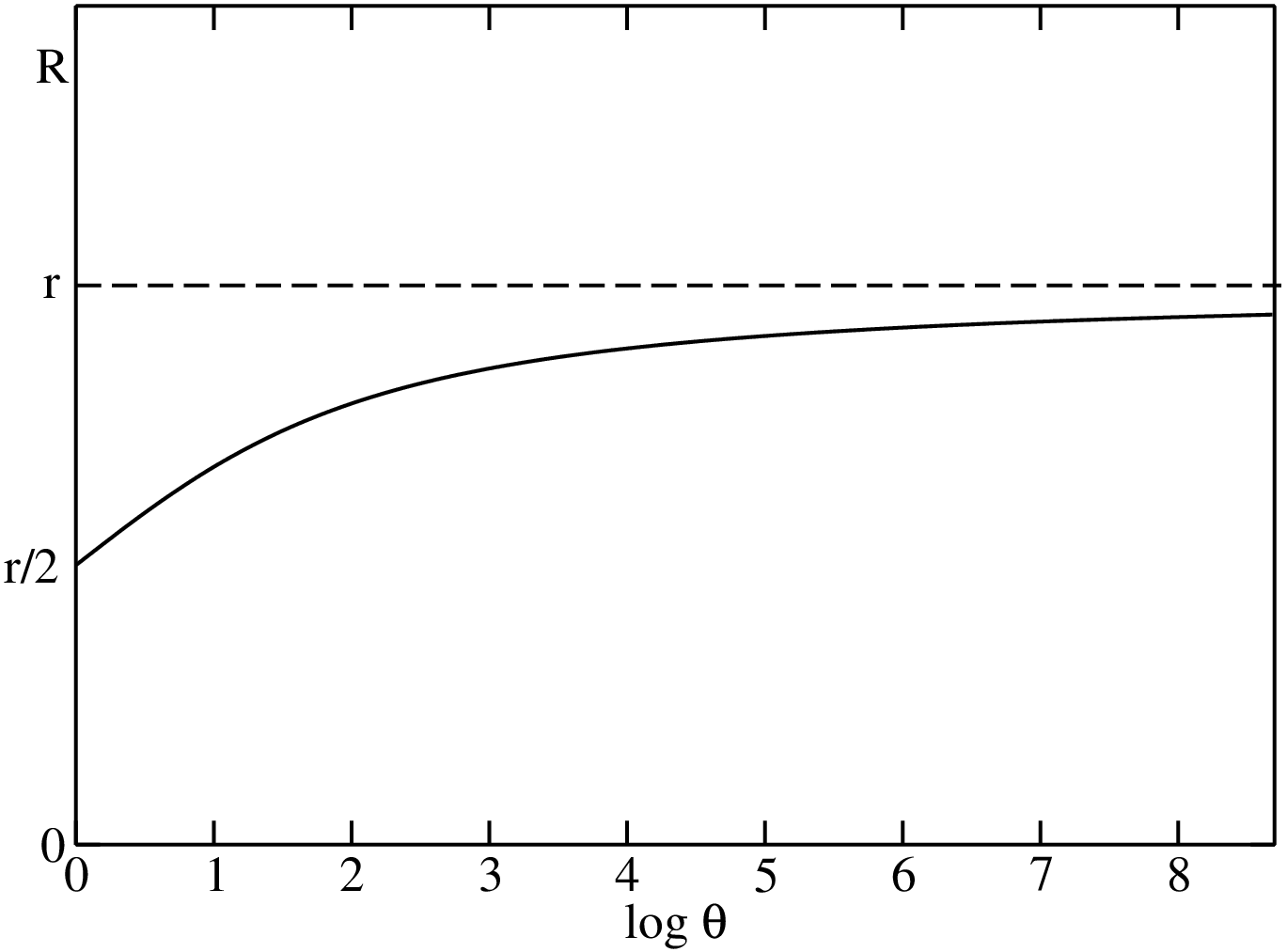}
\par\end{centering}
\caption{Function $R($$\theta)$ vs $\log\theta$ for $r=2/3$ (Eq. (\ref{9})).
\label{Fig.3}}
\end{figure}

The distribution given by Eq. (\ref{2}) indicates a change in the
parameter $\beta$ of the standard GR distribution. We denote by $B$
the modified parameter $\beta$; it is given by 
\begin{equation}
e^{-\beta M}\frac{2}{1+e^{-\beta M}}=e^{-BM}\,\,\,,\label{8}
\end{equation}
where $B$ is a function of $M$ ($B(M)$). It is convenient to introduce
the ratio $R=B/b$ (similar to $r=\beta/b$ given above), such that
Eq. (\ref{8}) becomes 
\begin{equation}
R=\frac{1}{\ln\theta}\ln\left[\frac{1}{2}\left(1+\theta^{r}\right)\right]\,\,\,,\label{9}
\end{equation}
where $\theta=\tau/\tau_{0}$ from Eq. (\ref{6}). The parameter $R$
varies from $R=r$ for large values of the variable $\theta$ to $R=r/2$
for $\theta\rightarrow1$ ($\tau\rightarrow\tau_{0}$). The function
$R(\theta)$ is plotted in Fig. \ref{Fig.3} \emph{vs} $\log\theta$
for $r=2/3$. The decrease of the function $R(\theta)$ for $\theta\longrightarrow1$
indicates correlations. 

According to Eq. (\ref{8}), the modified GR parameter $B$ is given
approximately by 
\begin{equation}
B(M)\simeq\beta-\frac{\ln2}{M}\,\,\,,\label{10}
\end{equation}
or 
\begin{equation}
R(\tau)\simeq r-\frac{\ln2}{\ln(\tau/\tau_{0})}\label{11}
\end{equation}
 for a reasonable range of foreshock magnitudes $M>1$. Eqs. (\ref{9})-(\ref{11})
show the decrease of the GR parameter in a foreshock sequence. For
instance, a $10\%$ decrease is achieved for $M=3$, or $\tau/\tau_{0}\simeq3.6\times10^{4}$
($\beta=2.3$, $r=2/3$). It is worth noting that smaller magnitudes
occur in the sequence of correlated foreshocks for shorter times,
measured from the occurrence of the main shock (the nearer main shock,
the smaller correlated foreshocks). 

On the other hand, the time-magnitude correlations expressed by Eq.
(\ref{6}) lead to $\tau=\tau_{0}e^{bM}$ for the accumulation time
elapsed from the main shock to an aftershock. This relation shows
a change in the seismicity conditions, where $t_{0}$ is replaced
by $\tau_{0}$ and $\beta$ is replaced by $b$ in the regular accumulation
time $t=t_{0}e^{\beta M}$. The magnitude distribution $\left(t_{0}/t^{2}\right)dt=\beta e^{-\beta M}dM$,
which follows from this accumulation time (Apostol 2006, see Appendix),
is changed in this case to $be^{-bM}dM$, which indicates an increase
in the GR parameter ($b=3.45$) with respect to its background value
$\beta$. Such a deviation holds up to a cutoff magnitude $M_{c}$
where the two distributions become equal, such that we may estimate
an average increase in the parameter $\beta$ as $(b-\beta)/2\beta=25\%$
for $\beta=2.3$. The cutoff magnitude is given by $be^{-bM_{c}}=\beta e^{-\beta M_{c}}$,
whence $M_{c}=0.36$ for $r=2/3$, $b=3.45$ ($\beta=2.3$). Both
these estimated deviations of the GR parameter for foreshocks and
aftershocks are in quantitative agreement with data reported by Gulia
et al. (2016, 2018) and Gulia and Wiemer (2019).
\begin{figure}
\begin{centering}
\includegraphics[clip,scale=0.3]{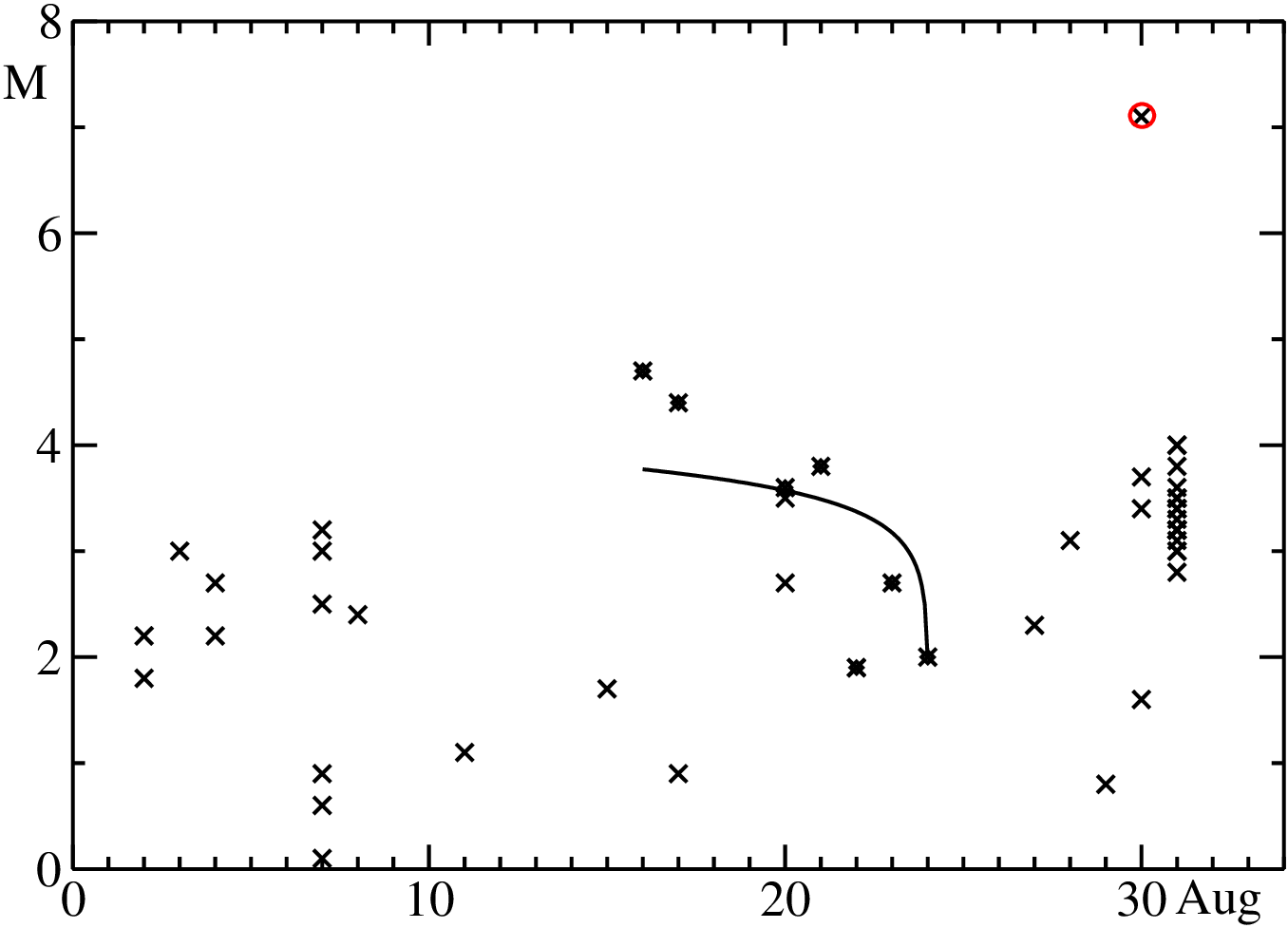}
\par\end{centering}
\caption{Vrancea seismic activity in the period 1 August - 31 August 1986 (Romanian
Earthquake Catalog, 2018). The curve is the fit of Eq. (\ref{12})
to data from 16 August to 24 August (fitting parameters $t_{ms}=24$
August and $\tau_{0}=10^{-4.76}$ days; see text).\label{Fig.4}}

\end{figure}

The logarithmic law expressed by Eq. (\ref{6}) for the time-magnitude
correlated foreshocks provides a means of estimating the occurrence
time of the main shock. Indeed, if we update the slope $B$ of the
cumulative distribution $\ln[N(M)/N(0)]=-BM$ at various successive
times $t$ (Eq. (\ref{8})), and if this $B$ fits Eq.  (\ref{10}),
then we may say that we are in the presence of a correlated sequence
of foreshocks which may announce a main shock at the moment $t_{ms}=t+\tau$.
(In particular, the probability of occurrence of a main shock with
magnitude $M_{0}$ increases in this case by a factor $\frac{\overline{B}}{\beta}e^{(\beta-\overline{B})M_{0}}$,
where $\overline{B}$ is the average value of the parameter $B$).
For practical use it is more convenient to use directly Eq. (\ref{6}),
which leads to the time dependence
\begin{equation}
M(t)=\frac{1}{b}\ln\frac{t_{ms}-t}{\tau_{0}}\label{12}
\end{equation}
of the foreshock magnitudes, for $(1-r)t_{ms}<t<t_{ms}-\tau_{0}$
($0<M<M_{0}$). This formula provides an estimate of the occurrence
moment of the main shock $t_{ms}$ from the correlated-foreshock magnitudes
$M(t)$ and the background seismicity parameter $\tau_{0}$; the occurrence
time is given by 
\begin{equation}
t_{ms}=t+\tau_{0}e^{bM(t)}\,\,.\label{13}
\end{equation}
It is worth noting that the time $t_{ms}$ depends on the magnitude
of the main shock, as expected ($M_{0}$, which enters $\tau_{0}$,
Eq. (\ref{7})). For instance, a magnitude $M$ indicates a time $\tau=\tau_{0}e^{bM}$
up to the main shock (Eq. (\ref{6})). Let us assume that we are interested
in a main shock with magnitude $M_{0}=7$; then by using $t_{0}=e^{-11.32}$
(years, for Vrancea) and $r=2/3$ given above, we get $\tau_{0}=\frac{2}{3}10^{-8.42}$
(years); a foreshock with magnitude $M=5$ would indicate that we
are at $\tau=\frac{2}{3}10^{-8.42}10^{7.5}=0.079$ years, \emph{i.e.}
$\simeq29$ days, from that main shock. The time $t_{ms}$ of the
occurrence of the main shock is obtained from Eq. (\ref{12}) as a
fitting parameter of the correlated-foreshock magnitudes $M(t)$.
In practice, it is also convenient to view $\tau_{0}$ as a fitting
parameter. Since, for moderate magnitudes, the variation of the parameter
$R$ is small (Eqs. (\ref{10}) and (\ref{11})), we may use the background
value for $r$ in the expression of $\tau_{0}$ (\emph{e.g.}, $r=2/3$),
which leads to an estimate of the expected main-shock magnitude $M_{0}$
from the fitting parameter $\tau_{0}$. However, a reliable estimation
of the time $t_{ms}$ provided by Eq. (\ref{13}) requires a very
high slope of the decreasing magnitudes $M(t)$ in the neighbourhood
of $t_{ms}$, which can only be attained by a special data set, including,
ideally, many small-magnitude foreshocks whose magnitudes fall rapidly
to zero. 

\section{Results}

The application of Eqs. (\ref{12}) and (\ref{13}) to fitting the
correlated foreshocks involves certain particularities. First, we
should note that not all the precursory seismic events are foreshocks
correlated with the main shock. Second, small clusters of precursory
events may exist, which may include second-order (an even higher-order)
correlated earthquakes, \emph{i.e.} events which accompany precursory
events, according to the epidemic-type model (see, for instance, Ogata
1988, 1998, as well as Helmstetter and Sornette 2003 and Saichev and
Sornette 2005). These secondary events have little relevance upon
a forthcoming main shock, such that they may be left aside. We limit
ourselves to the highest foreshocks occurring in short periods of
time (though an average magnitude for each small cluster may also
be used). Third, the relevant part of the logarithmic curve given
by Eq. (\ref{12}) (or the exponential in Eq. (\ref{13})) is its
abrupt decrease in the immediate proximity of $t_{ms}$ (of the order
of days for Vrancea), such that we should limit ourselves to foreshocks
which occur in the last few days. This limitation is related to the
very small values of the parameter $t_{0}$ and the large magnitude
$M_{0}$, of interest for the main shock (small values of the parameter
$\tau_{0}$). In this regard, a reliable estimation of the parameters
$t_{ms}$ and $\tau_{0}$ would be conditioned by a rich seismic activity
in the immediate vicinity of the occurrence moment of a main shock
(a very short-time prediction, \emph{e.g.}, of the order of days).
This is an ideal situation, which is not achieved in practice, since
the number of small-magnitude foreshocks is small in the immediate
vicinity of the main shock, precisely due to decrease of the parameter
$\beta$ ($B$, $R$). Therefore, such fits are necessarily of poor
quality. 

We give here a few examples of application of the fitting procedure
described above. 

Vrancea is the main seismic region of Romania. Reliable recordings
of earthquakes started in Romania around 1980. Since then, three major
earthquakes occurred in Vrancea: 30 August 1986, magnitude $M=7.1$;
30 May 1990, magnitude $M=6.9$; 31 May 1990, magnitude $M=6.4$ (Romanian
Earthquake Catalog 2018, http://www.infp.ro/data/romplus.txt). The
$7.1$-earthquake (depth $131km$) is shown in Fig. \ref{Fig.4},
together with all its precursory seismic events from 1 August to 31
August. All these earthquakes occurred in an area with dimensions
$\simeq100km\times80km$ ($45^{\circ}-46^{\circ}$ latitude, $26^{\circ}-27^{\circ}$
longitude), at various depths in the range $30km-170km$, except for
the events of 7-8 August and the $1.6$-event of 30 August, whose
depth was $5km-20km$. The subset of earthquakes from 16 August to
24 August can be fitted by Eq. (\ref{12}) with the fitting parameters
$t_{ms}=24$ August, $\tau_{0}=10^{-4.76}$ days and a large rms relative
error $0.32$. The maximum magnitude has been used for the earthquakes
which occurred in the same day, because, very likely, those with smaller
magnitude are secondary accompanying events of the greatest-magnitude
shock. If we assume that this is a correlated-foreshock subset, it
would indicate the occurence of a main shock with magnitude $4.4$
on 24 August. The main shock ($M=7.1$) occurred on 30 August. The
three earthquakes from 27 August to 29 August may belong to a subset
prone to such an analysis, but it is too poor to be useful. A main
shock with magnitude $7.1$ ($\tau_{0}=10^{-6.06}$ days) and an average
magnitude for the days with multiple events leads to a fit with a
larger rms relative error $0.6$. For the earthquake pair of 30-31
May 1990 (depth $87-91km$) we cannot identify a correlated subset
of foreshocks, \emph{i.e.} a sequence of precursory events with an
average magnitude, or a maximum-magnitude envelope, decreasing monotonously
in a reasonably short time range. Another particularity in this case,
in comparison to the earthquake of 1986, is the quick succession (30-31
May) of two comparable earthquakes (magnitude $6.9$-$6.4$). 

In the set of precursory events of the l'Aquila earthquake, 6 April
2009 (magnitude $6.3$, local magnitde $5.9$) one can identify two
magnitude-descending sequences, with earthquakes succeeding rapidly
at intervals of hours. The first sequence occurred on 2 April, consisting
of 7 earthquakes with local magnitudes from $2.1$ to $1.0$. The
fitting of these data with Eq. (\ref{12}) indicates a main shock
approximately 5 hours before the earthquake with magnitude $3.0$
of 3 April (with a large rms relative error $0.4$). The second magnitude-descending
sequence consists of 5 earthquakes with magnitudes from $1.9$ to
$1.1$, which ocurred on 6 April. The fit, with a similar large error,
indicates the occurrence of a main shock at the time $01:35$; the
l'Aquila earthquake occurred at $01:32$ (UTC; the last foreshock
was recorded at $01:20$). On the other hand, a magnitude-descending
sequence cannot be identified before the earthquake of 4 April, with
local magnitude $3.9$. The data used in this analysis are taken from
the Bollettino Sismico Italiano, 2002-2012, in $\pm25km$ an area
around the epicentre of the l'Aquila earthquake ($42.342^{\circ}$
latitude, $13.380^{\circ}$ longitude). The lack of the background
seismicity parameters $\beta$ and $-\ln t_{0}$ for the l'Aquila
region prevents us from estimating the magnitude of the main shocks
by this analysis. We note that the magnitude in the fitting Eq. (\ref{12})
is the moment magnitude; the use of local magnitudes in this Eq. generates
(small) errors. 

We applied the same procedure to the Vrancea earthquake with magnitude
$3.8$ (local magnitude $4.1$), viewed as a main shock, which occurred
on 30 November 2021 (Apostol and Cune 2021). By making use of the
foreshock sequence from 24 November to 27 November ($5$ earthquakes),
we can predict a main shock on 28 November, with a large magnitude
($6.9$, with a small rms relative error). On 28 November an earthquake
with magnitude $3.1$ was recorded in this area. By extending the
sequence until 29 November ($7$ earthquakes), a main shock with magnitude
$4.5$ was forecasted on 1 December (all the data are taken from Romanian
Earthquake Catalog 2018, http://www.infp.ro/data/romplus.txt). All
these earthquakes occurred within $45^{\circ}-46^{\circ}$ latitude,
$26^{\circ}-27^{\circ}$ longitude, at depths in the range $90km-180km$.

\section{Conclusions}

In conclusion, the GR distributions modified by correlations in the
foreshock region and the time dependence of the foreshock magnitudes
(Apostol 2021) can be used, in principle, to estimate the moment of
occurrence of the main shock and its magnitude, although with limitations.
The main source of errors arises from the quality of the fit $B(t)$
\emph{vs} $M(t)$ (Eq. (\ref{10})), or, equivalently, the fit of
the function $R(\theta)$ given by Eq. (\ref{9}), or the fit given
by Eqs. (\ref{12}) and (\ref{13}). These fits are necessarily of
a poor quality, due to the abrupt decrease of the function $M(t)$
near the occurrence time $t_{ms}$ of the main shock (equation (\ref{12})),
or, equivalently, the abrupt decrease of the parameters $B(M)$ and
$R(\tau)$ for small values of the variables $M$ and $\tau$. Another
source of errors arises from the background parameters $t_{0}$ and
$r$ ($\beta)$, which may affect considerably the exponentials in
the formula of the time cutoff $\tau_{0}$ (Eq. (\ref{7})). The procedure
described above is based on the assumption that the foreshock magnitudes
are ordered in time according to the law given by Eq. (\ref{6}).
However, according to the epidemic-type model, the time-ordered magnitudes
may be accompanied by smaller-magnitudes earthquakes, such that the
law given by Eq. (\ref{6}) may exhibit lower-side oscillations, and
the slope given by Eq. (\ref{11}) may exhibit upper-side oscillations.
Several subsets of correlated foreshocks may be identified (in accordance
with the epidemic-type model), as well as the absence of correlations.
In spite of all these limitations, a continuous monitoring of the
foreshock seismic activity by means of the procedure described in
this paper may give interesting information about a possible mainshock.
Also, the decrease of the GR parameter in the correlated foreshock
sequences and its increase in aftershock sequences, as identified
in the previous works (\emph{e.g.}, Gulia and Wiemer 2019), as well
as in the present one, is a valuable piece of information. 

\textbf{Acknowledgements}

The authors are indebted to the colleagues in the Institute of Earth\textquoteright s
Physics, Magurele-Bucharest, for many enlightening discussions. This
work was carried out within the Program Nucleu 2019, funded by Romanian
Ministry of Research and Innovation, Research Grant \#PN19-08-01-02/2019
and PN \#PN19-06-01-01/2019. Data used for the Vrancea region have
been extracted from the Romanian Earthquake Catalog, 2018. 

REFERENCES

Apostol, B. F. (2006). Model of Seismic Focus and Related Statistical
Distributions of Earthquakes. Phys. Lett. \textbf{A357} 462-466, \\
doi: 10.1016/j.physleta.2006.04.080. 

Apostol, B. F. (2019). An inverse problem in seismology: derivation
of the seismic source parameters from $P$ and $S$ seismic waves.
J. Seismol. \textbf{23} 1017-1030.

Apostol, B. F. (2021). Correlations and Bath's law. Results in Geophysical
Sciences. \textbf{5} 100011.

Apostol, B. F. \& Cune, L. C. (2021). Prediction of Vrancea Earthquake
of November 30 2021. Seism. Bull. \textbf{2}, Internal Report National
Institute for Earth's Physics, Magurele.

Bhattacharya, P., Chakrabarti, C. K., Kamal \& Samanta, K. D. (2009).
Fractal models of earthquake dynamics. Schuster, H. G., ed., \emph{Reviews
of Nolinear Dynamics and Complexity} pp.107-150. NY: Wiley. 

Bollettino Sismico Italiano (2002-2012), http://bolettinosismica.rm.ingv.it.

Dascher-Cousineau, K., Lay, T., Brodsky, E. E. (2020). Two foreshock
sequences post Gulia and Wiemer (2019). Seism. Res. Lett. \textbf{91}
2843-2850.

Dascher-Cousineau, K., Lay, T., Brodsky, E. E. (2021). Reply to \textquotedbl Comment
on 'Two foreshock sequences post Gulia and Wiemer (2019)' by K. Dascher-Cousineau,
T. Lay , and E. E. Brodsky\textquotedbl{} by L. Gulia and S. Wiemer.
Seism. Res. Lett. \textbf{92} 3259-3264.

De Santis, A., Cianchini, G., Favali, P., Beranzoli, L., Boschi, E.
(2011). The Gutenberg-Richter law and entropy of earthquakes: two
case studies in Central Italy. Bull. Sesim. Soc. Am. \textbf{101}
1386-1395.

Enescu, B., Struzik, Z., Kiyono, K. (2008). On the recurrence time
of earthquakes: insight from Vrancea (Romania) intermediate-depth
events. Geophys. J. Int. \textbf{172} 395-404.

Frohlich, C. \& Davis, S. D. (1993). Teleseismic $b$ values; or much
ado about $1.0$. J. Geophys. Res. \textbf{98} 631-644.

Gulia, L., Tormann, T., Wiemer, S., Herrmann, M., Seif, S. (2016).
Short-term probabilistic earthquake risk assessment considering time-dependent
b values. Geophys. Res. Lett. \textbf{43} 1100-1108.

Gulia, L., Rinaldi, A. P., Tormann, T., Vannucci, G., Enescu, B.,
Wiemer, S. (2018). The effect of a mainshock on the size distribution
of the aftershocks. Geophys. Res. Lett. \textbf{45} 13277-13287.

Gulia, L. \& Wiemer, S. (2019). Real-time discrimination of earthquake
foreshocks and aftershocks. Nature \textbf{574} 193-199.

Gulia, L. \& Wiemer, S. (2021). Comment on \textquotedbl Two foreshock
sequences post Gulia and Wiemer (2019)\textquotedbl{} by K. Dascher-Cousineau,
T. Lay, and E. E. Brodky. Seism. Res. Lett. \textbf{92} 3251-3258.

Helmstetter, A. \& Sornette, D., (2003). Foreshocks explained by cascades
of triggered seismicity. J. Geophys. Res.: Solid Earth \textbf{108},\\
https://doi.org/10.1029/2003JB002409.

Jones, L. M. (1994). Foreshocks, aftershocks and earthquake probabilities:
accounting for the Landers earthquake. Bull. Seism. Soc. Am. \textbf{84}
892-899.

Lay, T. \& Wallace, T. C. (1995). \emph{Modern Global Seismology}.
Academic Press, San Diego, California.

Lombardi, A. M. (2021). A normalized distance test for co-determining
completeness magnitude and $b$-values of earthquake catalogs. J Geophys.
Res.: Solid Earth \textbf{126} e2020 JB021242 https://doi.org/10.1029/2020JB021242.

Marzocchi, W., Spassiani, I., Stallone, A., Taroni, M. (2020). How
to be fooled for significat variations of the b-value. Geophys. J.
Int. \textbf{220} 1845-1856.

Ogata, Y. (1988). Statistical models for earthquakes occurrences and
residual analysis for point proceses. J. Amer. Statist. Assoc. \textbf{83}
9-27.

Ogata, Y. (1998). Space-time point-process models for earthquakes
occurrences. Ann. Inst. Statist. Math. \textbf{50} 379-402.

Ogata, Y. \& Tsuruoka, H. (2016). Statistical monitoring of aftershock
sequences: a case study of the 2015 $M_{w}$7.8 Gorkha, Nepal, earthquake.
Earth, Planets and Space \textbf{68} 44, 10.1186/s40623-016-0410-8.

Pelletier, J. D. (2000). Spring-block models of seismicity: review
and analysis of a structurally heterogeneous model coupled to the
viscous asthenosphere, Rundle, J. B., Turcote, D. L. \& Klein, W.,
eds. \emph{Geocomplexity and the Physics of Earthquakes}. vol. 120.
NY: Am. Geophys. Union. 

Romanian Earthquake Catalog (2018), http://www.infp.ro/data/romplus.txt,
10.7014/SA/RO.

Saichev, A. \& Sornette, D. (2005). Vere-Jones' self-similar branching
model. Phys. Rev. \textbf{E72} 056122.

Stein, S. \&, Wysession, M. (2003). \emph{An Introduction to Seismology,
Earthquakes, and Earth Structure,} Blackwell, NY.

Udias, A. (1999). \emph{Principles of Seismology}. Cambridge University
Press, NY.

\section{Appendix}

\subsection{Geometric-growth model of energy accumulation in focus}

In Apostol (2006) a typical earthquake is considered, with a small
focal region localized in the solid crust of the Earth. The dimension
of the focal region is so small in comparison to our distance scale,
that we may approximate the focal region by a point in an elastic
body. The movement of the tectonic plates may lead to energy accumulation
in this pointlike focus. The energy accumulation in the focus is governed
by the continuity equation (energy conservation)
\begin{equation}
\frac{\partial E}{\partial t}=-\boldsymbol{v}gradE\,\,\,,\label{A1}
\end{equation}
where $E$ is the energy, $t$ denotes the time and $\boldsymbol{v}$
is an accumulation velocity. For such a localized focus we may replace
the derivatives in equation (\ref{A1}) by ratios of small, finite
differences. For instance, we replace $\partial E/\partial x$ by
$\Delta E/\Delta x$, for the coordinate $x$. Moreover, we assume
that the energy is zero at the borders of the focus, such that $\Delta E=-E$,
where $E$ is the energy in the centre of the focus. Also, we assume
a uniform variation of the coordinates of the borders of this small
focal region, given by equations of the type $\Delta x=u_{x}t$, where
$\boldsymbol{u}$ is a small displacement velocity of the medium in
the focal region. The energy accumulated in the focus is gathered
from the outer region of the focus, as expected. With these assumptions
equation (\ref{A1}) becomes 
\begin{equation}
\frac{\partial E}{\partial t}=\left(\frac{v_{x}}{u_{x}}+\frac{v_{y}}{u_{y}}+\frac{v_{z}}{u_{z}}\right)\frac{E}{t}\,\,.\label{A2}
\end{equation}
Let us assume an isotropic motion without energy loss; then, the two
velocities are equal, $\boldsymbol{v}=\boldsymbol{u}$, and the bracket
in equation (\ref{A2}) acquires the value $3$. In the opposite limit,
we assume a one-dimensional motion. In this case the bracket in equation
(\ref{A2}) is equal to unity. A similar analysis holds for a two
dimensional accumulation process. In general, we may write equation
(\ref{A2}) as 
\begin{equation}
\frac{\partial E}{\partial t}=\frac{1}{r}\frac{E}{t}\,\,\,,\label{A3}
\end{equation}
where $r$ is an empirical (statistical) parameter; we expect it to
vary approximately in the range $(1/3,\,1)$. We note that equation
(\ref{A3}) is a non-linear relationship between $t$ and $E$. The
parameter $r$ may give an insight into the geometry of the focal
region. This is why we call this model a geometric-growth model of
energy accumulation in the focal region.

The integration of equation (\ref{A3}) needs a cutoff (threshold)
energy and a cutoff (threshold) time. During a short time $t_{0}$
a small energy $E_{0}$ is accumulated. In the next short interval
of time this energy may be lost, by a relaxation of the focal region.
Consequently, such processes are always present in a focal region,
although they may not lead to an energy accumulation in the focus.
We call them fundamental processes (or fundamental earthquakes, or
$E_{0}$-seismic events). It follows that we must include them in
the accumulation process, such that we measure the energy from $E_{0}$
and the time from $t_{0}$. The integration of equation (\ref{A3})
leads to the law of energy accumulation in the focus
\begin{equation}
t/t_{0}=(E/E_{0})^{r}\,\,.\label{A4}
\end{equation}
 The time $t$ in this equation is the time needed for accumulating
the energy $E$, which may be released in an earthquake (the accumulation
time). This is the time-energy accumulation eqaution referred to in
the main text. 

\subsection{Gutenberg-Richter law. Time probability}

The well-known Hanks-Kanamori law reads 
\begin{equation}
\ln\overline{M}=const+bM\,\,\,,\label{A5}
\end{equation}
 where $\overline{M}$ is the seismic moment, $M$ is the moment magnitude
and $b=3.45$ ($\frac{3}{2}$ for base $10$). In Apostol (2019) the
relation $\overline{M}=2\sqrt{2}E$ has been established, where $\overline{M}=\left(\sum_{ij}M_{ij}^{2}\right)^{1/2}$
(mean seismic moment), $M_{ij}$ is the tensor of the seismic moment
and $E$ is the energy of the earthquake. If we identify the mean
seismic moment with $\overline{M}$ in equation (\ref{A5}) we can
write
\begin{equation}
\ln E=const+bM\label{A6}
\end{equation}
 (another $const$), or 
\begin{equation}
E/E_{0}=e^{bM}\,\,\,,\label{A7}
\end{equation}
 where $E_{0}$ is a threshold energy (related to $const$). Making
use of equation (\ref{A4}), we get 
\begin{equation}
t=t_{0}e^{brM}=t_{0}e^{\beta M}\,\,\,,\label{A8}
\end{equation}
where $\beta=br$. From this equation we derive the useful relations
$dt=\beta t_{0}e^{\beta M}dM$, or $dt=\beta tdM$. If we assume that
the earthquakes are distributed according to the well-known Gutenberg-Richter
distribution, 
\begin{equation}
dP=\beta e^{-\beta M}dM\,\,\,,\label{A9}
\end{equation}
we get the time distribution 
\begin{equation}
dP=\beta\frac{t_{0}}{t}\frac{1}{\beta t}dt=\frac{t_{0}}{t^{2}}dt\,\,.\label{A10}
\end{equation}
 This law shows that the probability for an earthquake to occur between
$t$ and $t+dt$ is $\frac{t_{0}}{t^{2}}dt$; since the accumulation
time is $t$, the earthquake has an energy $E$ and a magnitude $M$
given by the above formulae (equations (\ref{A7}) and (\ref{A8})).
The law given by equation (\ref{A10}) is also derived (Apostol 2021)
from the definition of the probability of the fundamental $E_{0}$-seismic
events ($dP=-\frac{\partial}{\partial t}\frac{t_{0}}{t}dt$). We note
that this probability assumes independent earthquakes. This time probability
is referred to in the main text.

\subsection{Correlations. Time-magnitude correlations}

If two earthquakes are mutually affected by various conditions, and
such an influence is reflected in the above equations, we say that
they are correlated to each other. Also, we say that either one earthquake
is correlated with the other. Of course, multiple correlations may
exist, \emph{i.e.} correlations between three, four, etc earthquakes.
We limit ourselves to two-earthquake (pair) correlations. Very likely,
correlated earthquakes occur in the same seismic region and in relatively
short intervals of time. The physical causes of mutual influence of
two earthquakes are various. In Apostol (2021) three types of earthquake
correlations are identified. In one type the neighbouring focal regions
may share (exchange, transfer) energy. Since the energy accumulation
law is non-linear, this energy sharing affects the occurrence time.
We call these correlations time-magnitude correlations. They are a
particular type of dynamical correlations. In a second type of correlations
two earthquakes may share their accumulation time, which affects their
total energy. We call such correlations (purely) dynamical (or temporal)
correlations. Both these correlations affect the earthquake statistical
distributions; in this respect, they are also statistical correlations.
Finally, additional constraints on the statistical variables (\emph{e.g.},
the magnitude of the accompanying seismic event be smaller than the
magnitude of the main shock) give rise to purely statistical correlations.

Let an amount of energy $E$, which may be accumulated in time $t$,
be released by two successive earthquakes with energies $E_{1,2}$,
such as $E=E_{1}+E_{2}$ (energy sharing). According to the accumulation
law (equation (\ref{A4})) 
\begin{equation}
\begin{array}{c}
t/t_{0}=(E/E_{0})^{r}=\left(E_{1}/E_{0}+E_{2}/E_{0}\right)^{r}=\\
\\
=(E_{1}/E_{0})^{r}(1+E_{2}/E_{1})^{r}\,\,\,,
\end{array}\label{A11}
\end{equation}
 or
\begin{equation}
t=t_{1}\left[1+e^{b(M_{2}-M_{1})}\right]^{r}\,\,\,,\label{A12}
\end{equation}
 where $t_{1}=t_{0}(E_{1}/E_{0})^{r}$ is the accumulation time of
the earthquake with energy $E_{1}$ and magnitude $M_{1}$, and $M_{2}$
is the magnitude of the earthquake with energy $E_{2}$. From equation
(\ref{A12}) we get 
\begin{equation}
b(M_{2}-M_{1})=\ln\left[\left(1+\tau/t_{1}\right)^{1/r}-1\right]\,\,\,,\label{A13}
\end{equation}
where $t=t_{1}+\tau$, $\tau$ being the time elapsed from the occurrence
of the earthquake $1$ until the occurrence of the earthquake $2$.
If $\tau/t_{1}\ll1$, as in foreshock-main shock-aftershock sequences,
this equation gives, after some simple manipulations, 
\begin{equation}
M_{2}=\frac{1}{b}\ln\frac{\tau}{\tau_{0}}\,\,,\,\,\tau_{0}==rt_{0}e^{-b(1-r)M_{1}}\,\,.\label{A14}
\end{equation}
We can see that $\tau$ differs from the accumulation time of the
earthquake $2$ (compare to equation (\ref{A8})); it is given by
parameters which depend on the earthquake $1$ ($M_{1}$). If the
earthquake $1$ is viewed as a main shock, then the earthquake $2$
is a foreshock or an aftershock. These accompanying earthquakes are
correlated to the main shock (and the main shock is correlated to
them). Equations (\ref{A14}) are referred to in the main text as
the time-magnitude formula (\ref{6}).

\subsection{Correlations. Dynamical correlations}

Let us assume that an earthquake occurs in time $t_{1}$ and another
earthquake follows in time $t_{2}$. The total time is $t=t_{1}+t_{2}$,
so these earthquakes share their accumulation time, which affects
their total energy. These are (purely) dynamical (temporal) correlations.
According to equation (\ref{A10}) (and the definition of the probability),
the probability density of such an event is given by 
\begin{equation}
-\frac{\partial}{\partial t_{2}}\frac{t_{0}}{(t_{1}+t_{2})^{2}}=\frac{2t_{0}}{(t_{1}+t_{2})^{3}}\label{A15}
\end{equation}
 (where $t_{0}<t_{1}<+\infty$, $0<t_{2}<+\infty$). By passsing to
magnitude distributions ($t_{1,2}=t_{0}e^{\beta M_{1,2}}$), we get
\begin{equation}
d^{2}P=4\beta^{2}\frac{e^{\beta(M_{1}+M_{2})}}{\left(e^{\beta M_{1}}+e^{\beta M_{2}}\right)^{3}}dM_{1}dM_{2}\label{A16}
\end{equation}
(where $0<M_{1,2}<+\infty$, corresponding to $t_{0}<t_{1,2}<+\infty$,
which introduces a factor $2$ in equation (\ref{A15})). This formula
(which is a pair, bivariate statistical distribution) is established
in Apostol (2021). If we integrate this equation with respect to $M_{2}$,
we get the distribution of a correlated earthquake (marginal distribution)
\begin{equation}
dP=\beta e^{-\beta M_{1}}\frac{2}{\left(1+e^{-\beta M_{1}}\right)^{2}}dM_{1}\,\,;\label{A17}
\end{equation}
if we integrate further this distribution from $M_{1}=M$ to $+\infty$,
we get the correlated cumulative distribution 
\begin{equation}
P(M)=\int_{M}^{\infty}dP=e^{-\beta M}\frac{2}{1+e^{-\beta M}}\,\,.\label{A18}
\end{equation}
 equations (\ref{A17}) and (\ref{A18}) are referred to in the main
text (equations (\ref{1}) and (\ref{2})).

For $M\gg1$ the correlated distribution becomes $P(M)\simeq2e^{-\beta M}$
and $\ln P(M)\simeq\ln2-\beta M$, which shows that the slope $\beta$
of the logarithm of the independent cumulative distribution (Gutenberg-Richter,
standard distribution $e^{-\beta M}$) is not changed (for large magnitudes);
the correlated distribution is only shifted upwards by $\ln2$. On
the contrary, for small magnitudes ($M\ll1$) the slope of the correlated
distribution becomes $\beta/2$ ($P(M)\simeq1-\frac{1}{2}\beta M+...$
by a series expansion of equation (\ref{A18})), instead of the slope
$\beta$ of the Gutenberg-Richter distribution ($e^{-\beta M}\simeq1-\beta M+...$).
The correlations modify the slope of the Gutenberg-Richter standard
distribution for small magnitudes. This is the roll-off effect referred
to in the main text. 
\end{document}